\documentclass[twocolumn,amsmath,amssymb]{revtex4}
\usepackage[dvips]{graphicx}
\usepackage{dcolumn}
\usepackage{bm}

\begin{document}
\title[]
{Theoretical construction of stable traversable wormholes}
\author{Peter K.\,F. Kuhfittig}
\address{Department of Mathematics\\
Milwaukee School of Engineering\\
Milwaukee, Wisconsin 53202-3109 USA}

\begin{abstract}\noindent
It is shown in this paper that it is possible, at least in principle,
to construct a traversable wormhole that is stable to linearized
radial perturbations by specifying relatively simple conditions on
the shape and redshift functions.
\end{abstract}

\maketitle
\noindent
PAC numbers: 04.20.Jb, 04.20.Gz


\section{Introduction}\noindent
Wormholes may be defined as handles or tunnels in the spacetime
topology linking widely separated regions of our Universe or of
different universes altogether.  That such wormholes may be traversable
by humanoid travelers was first conjectured by Morris and Thorne \cite
{MT88} in 1988.  To hold a wormhole open, violations of certain energy
conditions must be tolerated.

Another frequently discussed topic is stability, that is, determining
whether a wormhole is stable when subjected to linearized perturbations
around a static solution.  Much of the earlier work concentrated on
thin-shell Schwarzschild wormholes using the cut-and-paste technique
\cite{PV95}.  In this paper we are more interested in constructing
wormhole solutions by matching an interior traversable wormhole
geometry with an exterior Schwarzschild vacuum solution and examining
the junction surface.  (For further discussion of this approach, see
Refs. \cite{LLO03, fL04a, fL06a, LC05, LL08, IL02}.)  A linearized
stability analysis of thin-shell wormholes with a cosmological
constant can be found in Ref. \cite{LC04}, while Ref. \cite {fL05}
discusses the stability of phantom wormholes.  In other, related,
studies, the Ellis drainhole was found to be unstable to non-linear
perturbations \cite {SH02} but stable to linear perturbations
\cite {kB98}.  According to Refs. \cite{GGS1, GGS2}, however, such
wormholes are actually unstable to both types of perturbations.

A rather different approach to stability analysis is presented in
Ref. {\cite{ES07}.  In Ref. \cite{KM05}, an example of a stable
traversable wormhole connecting two branes in the Randall-Sundrum
model is considered, while Ref. \cite{SZ08} discusses the instability
of scalar wormholes in a cosmological setting.

The purpose of this paper is to show that it is in principle possible
to construct a traversable wormhole that is stable to linearized
radial perturbations.  The conditions on the redshift and shape
functions at the junction surface are surprisingly simple.

\section{Traversable wormholes}\label{S:traversable}\noindent
Using units in which $c=G=1,$ the interior wormhole geometry is
given by the following metric
\cite{MT88}:
\begin{equation}\label{E:line1}
 ds^2=-e^{2\Phi(r)}dt^2+\frac{dr^2}{1-b(r)/r}+r^2(d\theta^2
     +\text{sin}^2\theta\,d\phi^2).
\end{equation}
The motivation for this idea is the Schwarzschild line element
\begin{equation}\label{E:line2}
   ds^2=-\left(1-\frac{2M}{r}\right)dt^2+\frac{dr^2}{1-2M/r}
   +r^2(d\theta^2+\text{sin}^2\theta\,d\phi^2).
\end{equation}
In Eq.~(\ref{E:line1}), $\Phi=\Phi(r)$ is referred to as the
\emph{redshift function}, which must be everywhere finite to prevent
an event horizon.  The function $b=b(r)$ is usually referred to as
the \emph{shape function}.  The minimum radius $r=r_0$ is the \emph
{throat} of the wormhole, where $b(r_0)=r_0$.  To hold a wormhole open,
the weak energy condition (WEC) must be violated.  (The WEC requires
the stress-energy tensor $T_{\alpha\beta}$ to obey $T_{\alpha\beta}
\mu^{\alpha}\mu^{\beta}\ge 0$ for all time-like vectors and, by
continuity, all null vectors.)  As a result, the shape function must
obey the additional flare-out condition $b'(r_0)<1$ \cite{MT88}.  For
$r>r_0$, we must have $b(r)<r$, while $\lim_{r \to \infty}b(r)/r=0$
(asymptotic flatness).  Well away from the throat both $\Phi$ and
$b$ need to be adjusted, as we will see.

The need to violate the WEC was first noted in Ref.~\cite{MT88}.  A
well-known mechanism for this violation is the Casimir effect.  Other
possibilities are phantom energy \cite{sS05} and Chaplygin traversable
wormholes \cite{fL06}.

Since the interior wormhole solution is to be matched with an exterior
Schwarzschild solution at the junction surface $r=a$, denoted by $S$,
our starting point is the Darmois-Israel formalism \cite{wI66, mV95}:
if $K_{ij}$ is the extrinsic curvature across $S$ (also known as
the second fundamental form), then the stress-energy tensor
$S^i_{\phantom{i}j}$ is given by the Lanczos equations:
\begin{equation}\label{E:Lanczos}
  S^i_{\phantom{i}j}=-\frac{1}{8\pi}\left([K^i_{\phantom{i}j}]
   -\delta^i_{\phantom{i}j}[K]\right),
\end{equation}
where $[X]=\lim_{r\to a+}X-\lim_{r\to a-}X=X^{+}-X^{-}.$  So
$[K_{ij}]=K^+_{ij}-K^-_{ij},$ which expresses the discontinuity in the
second fundamental form, and $[K]$ is the trace of
$[K^i_{\phantom{i}j}]$.

In terms of the energy-density $\sigma$ and the surface pressure
$\mathcal{P}$, $S^i_{\phantom{i}j}=\text{diag}(-\sigma, \mathcal{P},
 \mathcal{P}).$  The Lanczos equations now yield
\begin{equation}\label{E:stress1}
  \sigma=-\frac{1}{4\pi}[K^\theta_{\phantom{\theta}\theta}]
\end{equation}
and
\begin{equation}\label{E:stress2}
  \mathcal{P}=\frac{1}{8\pi}\left([K^\tau_{\phantom{\tau}\tau}]
    +[K^\theta_{\phantom{\theta}\theta}]\right).
\end{equation}

A dynamic analysis can be obtained by letting the radius $r=a$ be a
function of time, as in Ref.~\cite{PV95}. According to Lobo
\cite{fL05}, the components of the extrinsic curvature are given by
\begin{equation}\label{E:exterior1}
  K^{\tau+}_{\phantom{\tau}\tau}=\frac{\frac{M}{a^2}+\overset{..}{a}}
  {\sqrt{1-\frac{2M}{a}+\overset{.}{a}^2}},
\end{equation}
\begin{equation}\label{E:exterior2}
  K^{\tau-}_{\phantom{\tau}\tau}=\frac{\Phi'\left(1-\frac{b(a)}{a}
     +\overset{.}{a}^2\right)+\overset{..}{a}
    -\frac{\overset{.}{a}^2\left[b(a)-ab'(a)\right]}{2a[a-b(a)]}}
        {\sqrt{1-\frac{b(a)}{a}+\overset{.}{a}^2}},
\end{equation}
and
\begin{equation}\label{E:exterior3}
  K^{\theta+}_{\phantom{\theta}\theta}=\frac{1}{a}
    {\sqrt{1-\frac{2M}{a}+\overset{.}{a}^2}},
\end{equation}
\begin{equation}\label{E:exterior4}
  K^{\theta-}_{\phantom{\theta}\theta}=\frac{1}{a}
    {\sqrt{1-\frac{b(a)}{a}+\overset{.}{a}^2}}.
\end{equation}

For future use let us also obtain $\sigma'$: from
\begin{multline}\label{E:sigma}
\sigma=-\frac{1}{4\pi}(K^{\theta+}_{\phantom{\theta}\theta}-
       K^{\theta-}_{\phantom{\theta}\theta})=\\
   -\frac{1}{4\pi a}\left(\sqrt{1-\frac{2M}{a}+\overset{.}{a}^2}-
    \sqrt{1-\frac{b(a)}{a}+\overset{.}{a}^2}\right),
\end{multline}
one can calculate
\begin{multline}\label{E:sigmaprime}
  \sigma'=\frac{\overset{.}{\sigma}}{\overset{.}{a}}=
     \frac{1}{4\pi a^2}
   \left(\frac{1-\frac{3M}{a}+\overset{.}{a}^2-a\overset{..}{a}}
      {\sqrt{1-\frac{2M}{a}+\overset{.}{a}^2}}\right.\\
   \left.-\frac{1-\frac{3b(a)}{2a}+\frac{b'(a)}{2}
           +\overset{.}{a}^2-a\overset{..}{a}}
     {\sqrt{1-\frac{b(a)}{a}+\overset{.}{a}^2}}\right).
\end{multline}
Again following Lobo \cite{fL05}, rewriting Eq.~(\ref{E:sigma}) in
the form
\begin{equation}\label{E:modified}
  \sqrt{1-\frac{2M}{a}+\overset{.}{a}^2}=
      \sqrt{1-\frac{b(a)}{a}+\overset{.}{a}^2}-4\pi\sigma a
\end{equation}
will yield the following equation of motion:
\begin{equation}\label{E:motion}
   \overset{.}{a}^2+V(a)=0.
\end{equation}
Here $V(a)$ is the potential, which can be put into the following
convenient form:
\begin{equation}\label{E:potential1}
  V(a)=1-\frac{\frac{1}{2}b(a)+M}{a}-\frac{m^2_s}{4a^2}
   -\left(\frac{M-\frac{1}{2}b(a)}{m_s}\right)^2,
\end{equation}
where
$m_s=4\pi a^2\sigma$ is the
mass of the junction surface, which is a thin shell \cite{fL05}.
(If the surface stresses are zero, then the junction is referred
to as a boundary surface.)

When linearized around a static solution at $a=a_0$, the solution is
stable if, and only if, $V(a)$ has a local minimum value of zero at
$a=a_0$, that is, $V(a_0)=0$ and $V'(a_0)=0$, and its graph is
concave up: $V''(a_0)>0$.  For $V(a)$ in Eq.~(\ref{E:potential1}), these
conditions are met \cite{LC05}.

Since the junction surface $S$ is understood to be well away from
the throat, we expect $\sigma$ to be positive.  Eq.~(\ref{E:sigma}) then
implies that $b(a)<2M$, rather than $b(a)=2M$, which the Schwarzschild
line element (\ref{E:line2}) might suggest.  The reason is that the
interior and exterior regions may be separated by a thin shell, which
is not part of the interior solution.  (Also, in its most general form,
the junction formalism joins two distinct spacetime manifolds $M_+$ and $M_-$
with metrics given in terms of independently defined coordinate
systems $x_+^{\mu}$ and $x_-^{\mu}$ \cite{LC05}.)  What needs to be
emphasized is that even if $b(a) <2M$, $b(a)$ can be arbitrarily close
to $2M$ without affecting the above analysis.  In particular,
$V(a_0)=0$ and $V'(a_0)=0$ even if $\lim_{a\to a_0-}b(a)=2M$.
The condition $V''(a_0)>0$ should now be written $V''(a_0-)>0$.

\emph{Remark:} The assumption that $b(a)<2M$ is not actually
necessary.  If $\sigma$ is negative, then $2M$ is diminished,
so that $b(a)>2M$.  But we still have $\lim_{a\to a_0-}b(a)=2M$.

\section{The line element}\label{S:lineelement}\noindent
Given our aim, the construction of a stable wormhole, our main
requirement can now be stated as follows: apart from the usual
conditions at the throat, we require that $b=b(r)$ be an
increasing function of $r$ having a continuous second derivative
and reaching a maximum value at some $r=a$.  In other words, we
require that $b'(r)$ approach zero continuously as $r\rightarrow a$
(Fig. 1).  Keeping in mind the Schwarzschild line element
\begin{figure}[htbp]
\begin{center}
\includegraphics [clip=true, draft=false, bb= 0 0 305 190,
angle=0, width=4.5 in,
height=2.5 in, viewport=40 40 302 185]{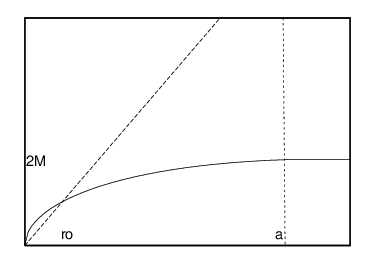}
\end{center}
\caption{\label{fig:figure1}The interior shape function attains
a maximum value at $r=a$.}
\end{figure}
(\ref{E:line2}), $b(r)$ becomes $b(a)=2M$ by continuity.  So by
Eq.~(\ref{E:sigma}), $\sigma=0$ at $r=a$.  It is also desirable
to have $\mathcal{P}=0$ at $r=a$, thereby making
$S^i_{\phantom{i}j}=0$.  To this end we choose $\Phi(r)$ so that
\begin{equation}\label{E:Phiprime}
  \Phi'(a-)=\frac{M}{a(a-2M)}.
\end{equation}
[Of course, $\Phi(r)$ must still be finite at the throat, while
$\Phi(a-)=\Phi(a+)$.]  For $r>a$, $\Phi(r)=\frac{1}{2}\,\text{ln}
(1-\frac{2M}{r})$, so that $\Phi'(a-)=\Phi'(a+)$. [One can also
say that $\Phi_{\text{internal}}(a)=\Phi_{\text{external}}(a)$
and $\Phi'_{\text{internal}}(a)=\Phi'_{\text{external}}(a)$.]
We now have
$K^{\tau+}_{\phantom{\tau}\tau}-K^{\tau-}_{\phantom{\tau}\tau}=0$
and $K^{\theta+}_{\phantom{\theta}\theta}-
K^{\theta-}_{\phantom{\theta}\theta}=0$ at $r=a$.  So $\mathcal{P}=0$
at $r=a$ by  Eqs. (\ref{E:stress2})-(\ref{E:exterior4}), as
desired.

For the above choice of $\Phi(r)$, the resulting line element is
\begin{multline*}
 ds^2=-e^{2\Phi(r)}dt^2+\frac{dr^2}{1-b(r)/r}+r^2(d\theta^2
     +\text{sin}^2\theta\,d\phi^2),\\ r\le a,
\end{multline*}
\begin{multline}\label{E:line3}
 ds^2=-e^{2\Phi(r)}dt^2+\frac{dr^2}{1-b(a)/r}+r^2(d\theta^2
     +\text{sin}^2\theta\,d\phi^2),\\ r>a.
\end{multline}
Note especially that
\begin{equation}\label{E:continuous}
  \frac{d}{dr}g_{tt}(a-)=\frac{d}{dr}g_{tt}(a+)\,\, \text{and}
\,\, \frac{d}{dr}g_{rr}(a-)=\frac{d}{dr}g_{rr}(a+).
\end{equation}
Since the components of the stress-energy tensor are equal to zero
at $S$, the junction is a boundary surface, rather than a thin shell~
\cite{fL05}, and $K_{ij}$ is continuous across $S$.

\section{Stability}\noindent
As noted at the end of Sec. \ref{S:traversable}, $V(a_0-)=0$ and
$V'(a_0-)=0$ even if $\lim_{a\to a_0-}b(a)=2M$.
In Sec.~\ref{S:lineelement} we saw that in the absence of surface
stresses our junction is a boundary surface, rather than a thin shell:
since $b'(r)$ goes to zero continuously as $r\rightarrow a-$, $b(r)$
continues smoothly at $r=a$ to become $2M$ to the right of $a$
(Fig. 1).  This implies that the usual thin-shell formalism using the
$\delta$-function is not directly applicable. To show this, suppose
we write the derivatives in Eq.~(\ref{E:continuous}) in the following
form:
\[
  \frac{d}{dr}g_{\mu\nu}=\Theta(r-a)\frac{d}{dr}g^+_{\mu\nu}(r)
     +\Phi[-(r-a)]g^-_{\mu\nu}(r),
\]
where $\Theta$ is the Heaviside step function.  Then by the product
rule,
\begin{multline*}
  \frac{d^2}{dr^2}g_{\mu\nu}(a\pm)=\\
  \Theta(r-a)\frac{d^2}{dr^2}g_{\mu\nu}(a+)+\Theta[-(r-a)]
    \frac{d^2}{dr^2}g_{\mu\nu}(a-)\\
     +\delta(r-a)\left(\frac{d}{dr}g_{\mu\nu}(a+)
         -\frac{d}{dr}g_{\mu\nu}(a-)\right).
\end{multline*}
So by Eq.~(\ref{E:continuous}).
\begin{multline*}
  \frac{d^2}{dr^2}g_{\mu\nu}(a\pm)=\\
  \Theta(r-a)\frac{d^2}{dr^2}g_{\mu\nu}(a+)+\Theta[-(r-a)]
    \frac{d^2}{dr^2}g_{\mu\nu}(a-).
\end{multline*}
Up to the second derivatives, then, the $\delta$-function does not
appear, in agreement with Visser~\cite{mV95}: by adopting Gaussian
normal coordinates, the total stress-energy tensor, which also
depends on the second derivatives of $g_{\mu\nu}$,  may be written
in the form $T_{\mu\nu}=\delta(\eta)S_{\mu\nu}
+\Theta(\eta)T^{+}_{\mu\nu}+\Theta(-\eta)T^{-}_{\mu\nu}$, thereby
showing the $\delta$-function contribution at the location of the
thin shell; here $\sigma$ is necessarily nonzero.  In our
situation, however, $S_{ij}=0$ at the boundary surface, so that, once
again, the $\delta$-function does not appear.

Even more critical in the stability analysis is the need to study the
second derivative of $V(a)$ in Eq.~(\ref{E:potential1}).  Since
$V''(a)$ involves $m''_s=(4\pi a^2\sigma)''$, let us first use
Eq.~(\ref{E:stress1}) to write $\sigma'$ in the following form:
\[
  \sigma'=-\frac{1}{4\pi}\left(\Theta(r-a)\frac{d}{dr}
  K^{\theta+}_{\phantom{\theta}\theta}+
  \Theta[-(r-a)]\frac{d}{dr}
    K^{\theta-}_{\phantom{\theta}\theta}\right).
\]
As long as $b(r)$ is an increasing function without the assumed
maximum value at $r=a$, $\sigma'$ will have a jump discontinuity at
$r=a$.  So $\sigma''$ is equal to $\delta(r-a)$ times the magnitude
of the jump \cite{lS50}. If $b'(a_0)=0$, on the other hand, the
calculations leading to Eq.~(\ref{E:sigmaprime}) show that $\sigma'$
is continuous at $a=a_0$.  It follows that there is no
$\delta$-function in the expression for $V''(a_0)$.

Without the $\delta$-function, one cannot simply
declare $4\pi a^2\sigma$ to be the (finite) mass of the spherical
surface $r=a$, since the thickness of an ordinary surface is
undefined.  (It is quite another matter to assert that $dm=
4\pi\sigma a^2da$, which can indeed be integrated over a
\emph{finite} interval.)

Returning to Eq.~(\ref{E:sigmaprime}), when $\sigma'$ is evaluated
at the static solution, then $b'(a_0)=0$ implies that
$\sigma'(a_0)=0$.  So if $\sigma$ is positive (resp. negative),
then $\sigma$ approaches zero, its minimum (resp. maximum) value,
continuously as $a\rightarrow a_0-$, and, as a consequence, $\sigma>0$
(resp. $\sigma<0$) in the open interval $(a_0-\epsilon,a_0)$; here
$\epsilon$ is arbitrarily small, but finite (as opposed to infinitesimal).
As a result, $\sigma$ is approximately constant, but nonzero, in the
boundary layer extending from $r=a_0-\epsilon$ to $r=a_0$.  So for
$\overline{a}\in (a_0-\epsilon,a_0)$, $m_s=4\pi \overline{a}^2\sigma$
is a positive (resp. negative) constant, but one that can be made
as small as we please in absolute value.  Referring
back to Eq.~(\ref{E:potential1}), we now find the second derivative
of $V$, making use of the condition $b'(a_0)=0$.  Since $m_s$ is
fixed, we get
\begin{multline}\label{E:concaveup}
  V''(a_0-)=-\frac{\frac{1}{2}b''(a_0-)}{a_0-}-\frac{b(a_0-)+2M}
     {(a_0-)^3}\\
     -\frac{3m^2_s}{2(a_0-)^4}
        +\frac{b''(a_0-)}{m^2_a}[M-\frac{1}{2}b(a_0-)].
\end{multline}
Since $m^2_s$ is arbitrarily small, but nonzero, the third term on
the right-hand side is arbitrarily close to zero, while the last
term is equal to zero, since $M=\frac{1}{2}b(a_0-)$.  From
$V''(a_0-)>0$, we obtain
\[
  b''(a_0-)<-\frac{2[b(a_0-)+2M]}{(a_0-)^2}.
\]
Using our arbitrary $\epsilon$, we can also say that
\[
  b''(a_0-\epsilon)<-\frac{2[b(a_0-\epsilon)+2M]}{(a_0-\epsilon)^2}.
\]
The continuity of $b''(r)$ and $a^2$ now implies that
\begin{equation}\label{E:criterion}
   b''(a_0)<-\frac{8M}{a_0^2}.
\end{equation}
This is the stability criterion.

\vspace{12pt}
\section{Discussion}\noindent
This paper discusses the stability of Morris-Thorne and other
traversable wormholes, each having the metric given by
Eq.~(\ref{E:line1}), where $\Phi(r)$ and $b(r)$ are the redshift
and shape functions,
respectively.  The shape function is assumed to satisfy the usual
flare-out conditions at the throat, while the redshift function is
assumed to be finite.  The interior traversable wormhole solution is
joined to an exterior Schwarzschild solution at the junction surface
$r=a$, where $\Phi$ and $b$ must meet the conditions discussed in
Sec. \ref{S:lineelement}.  Our main conclusion is that the wormhole
is stable to linearized radial perturbations if $b=b(r)$ satisfies
the following condition at the static solution $a=a_0$:
$b''(a_0)<-8M/a_0^2$, where $M$ is the total mass of the
wormhole in one asymptotic region.

Since the curve $b=b(r)$ is concave down, $b''(a_0)<0$, but its
curvature has to be sufficiently large in absolute value to overtake
$8M/a_0^2=4b(a_0)/a_0^2$.  This condition is simple enough to suggest
that the form of $b(r)$ can be easily adjusted ``by hand."

A function that meets the condition locally can also be obtained by
converting the above inequality to the differential equation
\[
   b''(r)+\frac{4b(r)}{r^2}=-\lambda,
\]
where $\lambda$ is a small positive constant.  Confining ourselves
to the interval $(a_1,a_0]$, a solution is
\[
   \overline{b}(r)=c\sqrt{r}\,\text{sin}\left(\frac{1}{2}
       \sqrt{15}\,\,\text{ln}\,r\right)-\lambda r^2.
\]
To the left of $a_1$, $\overline{b}(r)$ can be joined smoothly to a
function that meets the required conditions at the throat, thereby
completing the construction.

\end{document}